\documentclass[prl,onecolumn,showpacs,superscriptaddress,nofootinbib]{revtex4}
\usepackage{graphicx}
\usepackage{color}
\usepackage{mathrsfs}

\begin{document}


\title{CELL MECHANICS AND SIGNALISATION: SARS-CoV-2 HIJACKS  MEMBRANE LIQUID CRYSTALS AND CYTOSKELETAL FRACTAL TOPOLOGY}

\author{Christiane BINOT}
\affiliation{ 17410 Saint Martin de R\'e, France}

\author{Jean-Fran\c cois SADOC}
\email{jean-francois.sadoc@universite-paris-saclay.fr}
\affiliation{Laboratoire de Physique des Solides (CNRS-UMR 8502), B{\^a}t. 510, Universit{\'e} Paris-Saclay, 91405 Orsay cedex, France}

\author{Claude-Henri CHOUARD}
\affiliation{Acad\'emie nationale de m\'edecine 16 rue Bonaparte 75006 Paris France}

\begin{abstract}
We highlight changes to cell signaling under virus  invasion (with the example of SARS-CoV-2 ), involving disturbance of membranes, (plasma, mitochondrial, endothelial-alveolar) and of nanodomains, modulated by the cytoskeleton. Virus alters the mechanical properties of the membranes, impairing mesophase structures mediated by the fractal architecture initiated by actomyosin. It changes the topology of the membrane and its lipid composition distribution.
Mechanotransduction, self-organization and topology far from equilibrium are omnipresent.
We propose that the actomyosin contractility generates the cytoskeleton’s fractal organization. \\
We focus on three membranar processus:\\
-The transition from lamellar configuration in cell and viral membranes to a bi-continuous organization in the presence of ethanolamine; (The  energy for this transition is provided by change of the folding of the viral fusion protein from metastable to stable state).\\
-The action of  mitochondrial antiviral signaling protein  on the external mitochondrial envelope in contact with  mitochondrial-associated membranes,  modified by viral endoribonuclease, distorting innate immune response.\\
-The increased permeability of the epithelial-alveolar-pulmonary barrier involves the cytoskeleton membranes. The pulmonary surfactant, is also perturbed in its liquid crystal state.\\
Viral subversion disorganizes membrane structure and functions and thus the metabolism of the cell.
 We advocate systematic multidisciplinary exploration of membrane mesophases and their links with fractal dynamics, to enable novel therapies for SARS-CoV-2 infection.
 
\end{abstract}


\maketitle


%

%
%


\section{Introduction }

Cells behavior depends of a multitude of mechanisms and organisations which need to be interpreted through a multidisciplinary approach. Even if most of individual mechanisms are now well understood, their collective aspect is less clearly analyzed. For that purpose, in a first stage we will focus on the membrane structural complexity in interaction with the cytoskeleton. The membrane structure can be analyzed using the concept of liquid crystal, but also of all perturbations of this order mixing knowledge from physics, chemistry, biology,  mathematics, topology... The cytoskeleton has a hierarchical structure giving to it spatial and temporal organisation a fractal like aspect, which also mix physics, chemistry, biology,  mathematics... The cytoskeleton interacts with membranes, but also notably with nucleus and mitochondria, allowing a bridge between
extracellular matrix to  intracytoplasmic nucleus, for instance with integrins and cadherins\cite{WangTytell}.
This collective aspect marks a signaling path allowing information exchange between all these domains.

Etienne et al\cite{EtienneFouchard} emphasized these processes  seen in mechano-transduction. Cells convert mechanical energy into a biochemical response, and conversely convert biochemical stimuli into mechanical signals. As underlined by Na et al\cite{NaCollin}, rapid signal transduction within living cells and long-distance propagation throughout the organism are the basic specificities of mechano-transduction. The range of time-scales involved, presents fractal dynamics.
These mechanisms go far beyond molecular diffusion and transduction induced by soluble ligands. The spatio-temporal properties of actin cytoskeleton structures cover scales ranging from $10^{-3}$ to $10^{-6}$ seconds and $10^{-8}$ to $10^{-4}$ meters\cite{GuoLi}.

The concept of “active matter” developed by Joanny\cite{Joanny2006} opens up the study of emergent self-organization in matter far from equilibrium, using energy from ATP hydrolysis. Such is notably the case for the molecular motor of actomyosin.

An interesting example of the collective interactions of the membrane and the cytoskeleton mechanical properties is observed in viral infections.
We propose the  following hypothesizes
related to interaction between RNA virus and cells:\\
-Actomyosin contractility induces fractal organization of molecules in membrane and cytoplasm, regulating the equilibrium of cell dynamics so the relation between structure and function.\\
-Cell membrane and cytoplasm disturbances induced by viral invasion leads to multi-scale dysfunction, from molecule to organism.\\
-Cell signalling results from a signal integrating inseparable interactions between mechanical factors and the cell’s electrochemical and genetic information.

Focussing on the Severe Acute Respiratory Syndrome Corona Virus-2 (SARS-CoV-2] case, the virus hijacks cell signaling from extracellular matrix to  intracytoplasmic nucleus.
Moreover fusion between viral and host-cell membranes exploits the cell’s mechanical properties. There is a process  shared by all enveloped viruses, based on the physics of viral and human phospholipid membranes. It has notably been described for corona viruses, influenza, Ebola, Dengue and Nile viruses\cite{RichardZhang}.

Despite great structural diversity, there is a shared transformation of viral fusion protein conformation mediating fusion, enabling the virus to acquire a prior state apt for membrane fusion as confirmed in 2020 by Benhaim et al\cite{BenhaimLee} and Chakraborty et al\cite{ChakrabortyBhattacharjya}. This state involves acquiring specific mechanical properties in viral intermembrane zones. It has been shown that phosphatidylethanolamine (PE) plays a major role in the topological restructuration connecting two lamellar layers by a tunnel recalling locally so called bi-continuous cubic phases. The reference to cubic phases is an easy labeling term but is not physically significant as between the cellular and virus membrane there is only one tunnel and not a large number of interacting tunnels\cite{ClercLevelut}.

Innate immune response is another example of the involvement of membranes and thus of the cytoskeleton in SARS-CoV-2 pathogenesis, in the mitochondrial-associated membranes (MAM) studied by Jacob et al\cite{JacobsCoyne} and Horner et al\cite{HornerLiu}. The mitochondrial antiviral signaling protein (MAVS)  lies in the outer membrane of the mitochondria and is activated by the virus. MAVS becomes associated to the viral RNA, which in turn activates the nuclear factor kappa B (NF-$\kappa$B) signaling pathway, inducing inflammatory response by production notably of type-1 interferon (IFN-1). Now, the mitochondrial membrane is considered as an immune synapse and inflammation regulator.  Hackbart et al\cite{HacbartDeng} highlighted a mechanism used by the SARS-CoV-2 virus disturbing recognition of its RNA by host immune receptors.

Acute respiratory distress syndrome (ARDS) also shows the involvement of liquid crystals and fractal dynamics in viral infection\cite{AstutiYsrafil}. Lung elastic properties are impaired, pulmonary volume is reduced, and epithelial-alveolar lesions maintain the edema resulting from endothelial dysfunction. There is a direct correlation between reinforcement of cytoskeleton-mediated adhesion sites and increased substrate rigidity. Cytoskeletal remodeling leads to redistribution of intra- and inter-cellular adhesion, increasing endothelial permeability. Finally, alveolar physiology is directly related to pulmonary surfactant, which shows liquid crystal order\cite{Mitov}.

The present article comprises two parts: one concerning the physiological properties of contractile actomyosin, and one highlighting the hijacking of actomyosin functions by SARS-CoV-2.
But more generally it seems important to recall a general behavior unifying membrane processes under pathogen invasion, also including early neurodegenerative, oncogenic or prionic processus and more generally metabolic phenomena. This broadened our understanding over all the spatiotemporal scales\cite{ChouardBinotNeur,ChouardBinotHelyon,ChouardBinot2019} of the fundamental mechanisms of cell life, which the virus hijacks as soon as it binds to the target for angiotensin-converting enzyme 2 (ACE2) receptor. The virus alters the inseparable complex of mechanical, electrochemical, genetic and epigenetic signaling. Coronaviruses and other viruses, including influenza A, enter the cell through membrane domains enriched in tetraspanins as observed by Earnest et al\cite{EarnestHantak}. SARS-CoV-2 uses these domain gathered platforms with liquid crystal order at a scale of hundreds of nanometers.

Finally, our understanding of the processes leading to SARS-CoV-2 infection depends on both membrane liquid crystal state of phospholipids and the fractal properties of the cytoskeleton. In biology only few studies are related to these domains, despite numerous studies of the physical properties of human cells in recent decades. Living tissues are active multi-functional forms of matter. They self-organize by generating, detecting and responding to mechanical stress and then activate membrane signaling pathways, mediated by the dynamic activity of the molecular motor actomyosin as it has been underlined by several teams\cite{AhmedBetz,KosterHusain,AgarwalZaidel,HeadPatel}.

It seems important\cite{KechagiaIvaska} to emphasized on the transmission of information from the nucleus to the extracellular matrix and other cells, or vice-versa, via patterns of mechanical tension: e.g., interaction with ion channels mobilizing electrical charge and promoting chemical reactions. The amphiphilic phospholipids of membrane bilayers in lamellar phase show the properties of mesophases as a liquid crystal states of matter. The heterogeneity of the membrane structure modulated by actomyosin plays a prime role in this exchange of information. Gmachowski\cite{Gmachowski}  emphasized  on the importance of the fractal aspect observed in membrane compartmentalization. We attribute this aspect to
the contractility of cytoskeletal actomyosin. Coupling between membrane dynamics and active cytoskeletal processes governs molecular diffusion behavior. Plasma membrane compartmentalization by the underlying cytoskeleton triggers molecule diffusion inside the membrane (they escape to classical Brownian motion law). A therapeutic target have to take into account the interaction coupling membrane and cytoskeleton.

\section{Role of Actomyosin in membrane and cytoskeleton interactions }

\subsection{Diffusion process }

The contractility of cytoskeletal actomyosin mediates the following processes in which we consider the involvement of membrane mesophases and fractal dynamics:   the basics of cell signaling.

At nanometric scale, the plasma membrane presents a hierarchical aspect within which many membrane proteins show abnormal diffusion patterns of under-diffusion or diffusion blocked in a fractal space.
Recent studies by Sadegh et al\cite{SadeghHiggins} showed the plasma membrane to be hierarchically compartmentalized by a fractal of dynamic cortical actin. The membrane/cytoskeleton interface has a multi-scale fractal structure, with active cytoskeleton processes coupled to membrane dynamics. So it is important to consider that membrane molecule movements  depend on actomyosin activity.
Confinement and segregation of membrane components enable functional domains to form. Protein crowded membranes show a non-Gaussian multifractal effect and heterogeneous lateral spatiotemporal diffusion at nano- to micro-second scales as indicated by Krapf\cite{Krapf2018} and Jeon et al\cite{JeonMartinez} in 2018. Lipid and protein trajectories, detected on analysis of spatial variations in diffusion in the membrane plane, show clear alteration in membrane dynamics.
%
%
\begin{figure}[tbp]
\resizebox{1.0\textwidth}{!}{%
\includegraphics{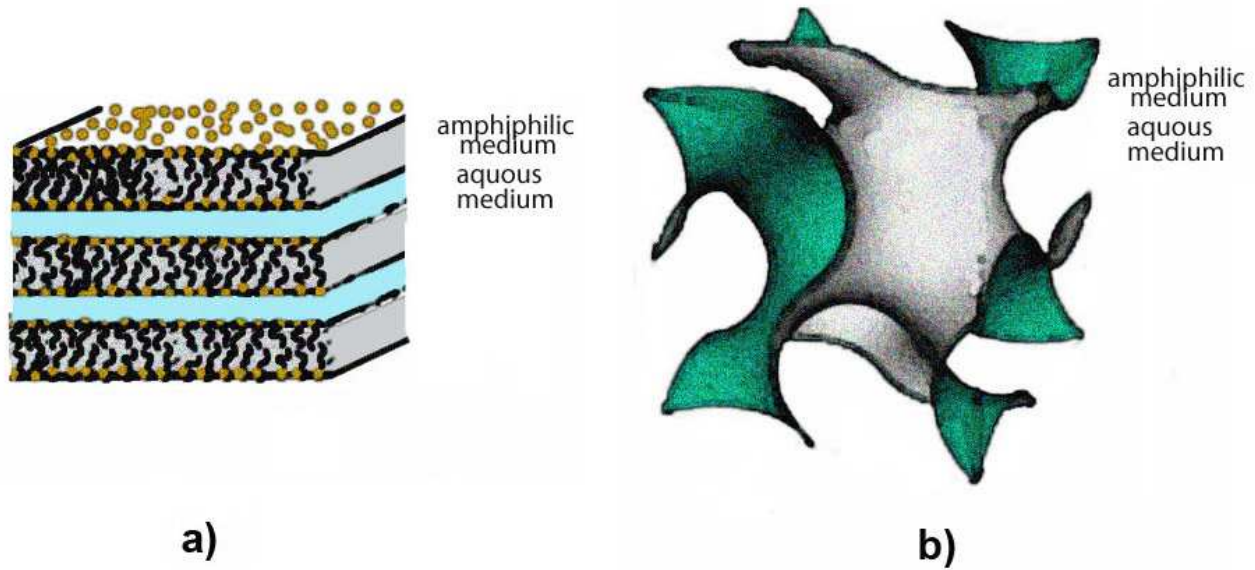} 
}
\caption{ a) Schematic representation of the lamellar
phase of a lyotropic liquid crystal,
amphiphilic flat layers  are separated by water layers ;
b) schematic representation of a cubic phase (gyroid surface), two interwoven but not connected labyrinths
of water channels are separated by a single film of  amphiphilic molecules with an intricate topology. In the lamellar phase layers have a null curvature when the gyroid surface has a negative Gaussian curvature. }
\label{f1}
\end{figure}
%
\subsection{Membranes and liquid crystal properties }
With a simplified point of view, the cell membrane is composed of various molecules, generically phospholipids, which under biological conditions interact fairly freely, interaction energies being of the same order as the thermal activation energy ($kT$ Boltzmann constant times absolute temperature) at living-tissue temperatures. See for instance temperatures for phase transitions measured in vitro experiments with molecules of biological origin reported in papers by Siegel et al.\cite{Siegel1997,Siegel1999} which are in the range $30 ^\circ C - 60 ^\circ C$; see also\cite{Chretien}.  In this temperature range, molecules like the phospholipids inter-diffuse freely, resulting in a liquid-like behavior: a given molecule $a$ neighboring an other molecule $b$ at time $t$ is unlikely to be still in the neighborhood after a short interval $dt$, even if mean distances between molecules remain very similar. There is few order binding molecules together. Even so, structures emerge involving large sets of molecules, which thus form a “liquid crystal”. This order, like that of classical crystals, may have very diverse forms, giving rise to the whole range of mesophases. In the case of membranes, a film-like structure (mathematically, 2-dimensional) emerges, organized in a bilayer which is the indication of their liquid crystals state probably associated to interaction of molecules located in both sides of the membrane.

Membranes are not pure liquid crystals, as they include a large variety of molecules of varying type, such as proteins or cholesterol for instance. Depending on the conditions, they may group together in a variety of ways, giving rise to islet of specific composition within the membrane: lipid rafts, or nanoscopic membrane domains with mesophase-like order. It is an important point to notice: membranes have a liquid crystals order underlying their structure but the discrepancies to this order leads to several aspects of their properties. As usual in physics, defects could be more important for the properties than the order itself.

Lipid bilayers are guided around an hypothetical mathematical surface. This is the amphiphilic nature of molecules with hydrophobic heads in contact with the cytoplasm or the extracellular environment, whereas within the film they have paraffinic fatty acids chains. The relative sectional sizes of the heads and paraffinic  tails determine film curvature: when equal, the film is flat, whereas small heads induce spherical bump with positive Gaussian curvature and oppositely large heads induce saddle like shape with negative Gaussian curvature.

As the membrane contains mixtures of molecules with differing characteristics, fluctuations in concentration can induce fluctuations in curvature, and vice-versa. Thus, the cytoskeleton may initiate fluctuations inducing a variety of compositions, deforming the membrane.

\subsection{Actomyosin and fractal behaviour}
Scalar properties show that non-linear regulation processes operate far from equilibrium and that regular processes are not necessarily dependent on maintained physiological control. They cover scales ranging from the molecular to the whole organism as studied by Goldberger et al\cite{GoldbergerLamaral}. Statistical physics reveals long-range correlations and power laws associated with multifractal cascades over a wide range of time-scales. The very rigorous mathematical concept of fractal structures can be applied to physical and biological objects analogous to mathematical objects. “Self-similarity” may apply over several ranges of scale; this scale-invariance encounters upper and lower limits in both time and space.

We suggest that these processes are due to actomyosin contractility generating fractal structures. To emphasize on fractal structures role, recall this example: dysfunction or loss of fractal organization and of non-linear interactions can cause sudden death by irreversible cardiac injury. Might not cardiac complications in viral infection be due to fractal order being disturbed by the viral invasion?

\subsection{Actomyosin, liquid ordered membrane domains and viral infection}
In 2017 Vogel et al\cite{VogelGreiss} propose the implication of a dynamic molecular organization in the membrane, in the form of rafts initiated by lateral diffusion of single molecules in membranes reconstituted at nanometric scale. Liquid ordered (Lo) phase domains could also be observed, as well as molecule diffuse out of the nanodomain. Vogel et al\cite{VogelGreiss} showed that, in a composite in-vitro layer, local membrane component organization was determined by actomyosin and its structure which can be a therapeutic aim.

The system associating actomyosin and plasma membrane is working as an active composite by continuous consumption of ATP; the remodeling is permanent. In the phase-segregated membrane bilayer, the remodeling actomyosin layer is indispensable for modifying the size and dynamics of the Lo domains, reducing or blocking their growth.
Many pathogens, including SARS-CoV-2, have been shown to use membrane domains to infect their target cell\cite{FavardChojnacki,Harrison2015}. We emphasize the role played by membrane location of the early processes involved in viral infection. It is within the membrane phospholipid bilayer, with its liquid crystal order, mediated biochemically by cholesterol and mechanically by the underlying cytoskeleton, that the virus triggers the first reaction to infection. We show how fusion peptides trigger a reorganization of membrane structure and composition following Vinson\cite{Vinson2020} and Florin et al\cite{FlorinThorsten}.

Thus the virus diverts cell signaling by modifying membrane molecular diffusion modulated by fractal organization of the cytoskeleton.  This fractal organization notably controls membrane domains. In this context it is primordial to take into account geometrical and topological properties of the membrane.

\subsection{Actomyosin and cell signaling}

 Cell signaling is directly related to plasma membrane compartmentalization\cite{TenchovMacdonald}, which shows a lateral organization in specific nanodomains, rearranged by fractal dynamics. We see this as the action of actomyosin modulating membrane domains and thereby controlling cell signaling. The SARS-CoV-2 virus uses tetraspanin-enriched membrane domains to anchor in the target cell.
The contractility of the actomyosin is an essential property, and it follows that it is this mechanical aspect of the association between membrane mesophases and cytoskeleton which has to be explored to understand their involvement in viral infection. The induction of stresses by actin and myosin, which is relevant of the processes of active mechanics, need to be taken into account in the aim of properly understanding all specificities of the membrane in living cells.

\subsection{Actomyosin and membrane fusion }

SARS-CoV-2 uses  mesophases containing PE to fuse its envelope with the host cell membrane phospholipids. This hypothesis is supported by experimental studies in physics showing that PE is directly involved in membrane restructuring, leading to fusion between viral and host cell membranes\cite{Shyamsunder1988}.
As suggested by Bulpett\cite{BulpettSnow}, in phospholipid system a transition from a lamellar to an inverted hexagonal order is energetically analogous to the membrane fusion process, in terms of curvature, elasticity, hydrophobicity, restructuring, etc. Knowledge of the physical behavior of liquid crystals can be applied to cell and viral membranes. The phospholipids which are their main constituent form a lipid bilayer, which can be analyzed as a liquid crystal. Viral entry involves a step of fusion between the cell and viral membranes\cite{Shyamsunder1988}, which, in terms of structural physics, amounts to a local topological change in liquid crystal structure\cite{XuNagy}. This involves diffusion of membrane molecules, like in a fractal medium, under the mechanical control of the intracellular cytoskeleton.

\begin{figure}[tbp]
\resizebox{1.0\textwidth}{!}{%
\includegraphics{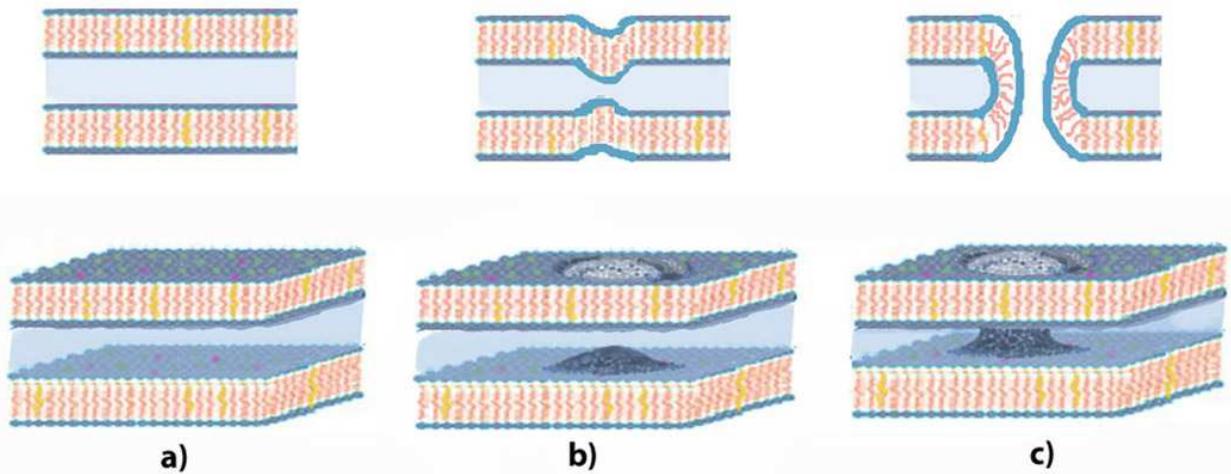}  
}
\caption{(a) Two bilayers of amphiphilic membranes. (b) Local change in properties of the phospholipid molecules create two bumps on the layers. (c) Fusion of the two layers leading to a bi-continuous structure. On (a) we can consider that there is three mediums : two phospholipidic membranes and a water part in sandwiches : this is a topological structuration. In (b) the local deformation keeps this topology unchanged. But in (c) the two membranes fusion changes the topology : there is still only one phospholipidic medium. On top is a very simplified scheme of the transformation.}
\label{f3}
\end{figure}

Conformal change occurring in membranes during fusion are modification of the topology of their organization. As usual topological transformations are characterized by change in medium connectivity (aquous or phospholipidic domain for instance; see figure~\ref{f3}). They enter in a large field of investigation in physics including for instance topological defects which are the local expression of spatial heterogeneity of an order, and play a determining role in the physical properties of matter. They can also be seen within active matter, a state characterized by large-scale mass flows emerging from self-organization of interacting self-propelled particles. Abnormal large fluctuations in density characterize active systems, from which dynamic turbulences emerge from active particles.

Saw\cite{SawDoostmohammadi} showed that epithelial topological defects govern cell death and extrusion. These active factors notably entail polymerization of actomyosin and cell contractility involving cell viscosity and elasticity. These developments should be included in the study of SARS-CoV-2 infection action mechanisms.

\subsection{Actomyosin and cell adhesion}
We emphasize the mechanical importance of cell adhesions, able to form immune synapses with the help of tetraspanin proteins. We consider that the virus exploits cellular mechanical-chemical principles, triggering immune responses that overwhelm the organism’s defenses\cite{}.
Adhesions, such as cell-cell adhesion, notably via cadherin proteins, and cell-matrix adhesion, via integrin proteins, respond to mechanical changes in the extracellular matrix, especially regarding rigidity. They modulate mechanical signaling in reaction to micro-environmental factors. Integrin proteins ensure cell adhesion to the extracellular matrix scaffolding and notably initiate local signaling within cell contacts and also remotely toward the cytoplasm and nucleus.

\subsection{Actomyosin, local order in membranes}

Some liquid crystal molecules self-associate to form 2D or 3D periodic structures at a scale of a few tens of nanometers. So-called “bi-continuous” cubic phases, for example, comprise two nested labyrinths separated by a periodic surface. These structures extend over much greater scales than those of the organizations described here. Even so, their local order results from interactions within distances between a few tens or a few hundreds of the constituent molecules. These are the interactions found in biological membranes and viral envelopes. Analysis of amphiphilic molecule liquid crystals thus sheds light on the mechanisms at work.

The classical crystallography of periodic organizations such as atoms or molecules are topologically considered as packing of objects without spatial dimensions (point like). But, more than thirty years ago, J. Charvolin and J.F. Sadoc have emphasized on the description of amphiphilic molecules constituting 3D liquid crystal structures that go beyond the classical description. Amphiphilic molecules assemble into the form of a surface, so an organization of two-dimensional films. The order inside the films is that of a liquid, while the surfaces are organized with a periodic order of crystals: this is a pure liquid crystal\cite{CharvolinSadoc1987,CharvolinSadocPolSci1990,CharvolinSadocJPhys1990,CharvolinSadoc2011}.

Understanding the local order in these systems required a geometric approach conciliating often incompatible topologic factors such as local film curvature and spatial extension. In this context, J. Charvolin and J.F. Sadoc\cite{SadocCharvolin2009} showed that bi-continuous or cellular topologies are relevant of ordered geometric configurations, analyzed in terms of topologic defects.

\subsection{Geometric approach to membrane fusion, involvement of liquid crystals and cytoskeleton dynamic activity}

Membrane fusion is a classical problem in soft matter physics when transition between lamellar phases and
other phases formed by phospholipidic molecules are observed and modeled\cite{CharvolinSadoc1996}.
We consider these structures as periodic systems of fluid film separated by interfaces, distinguishing lamellar phases, consisting of periodic layering and flat interfaces with constant distances, and cubic or hexagonal phases, symmetrically curved and layered periodically.

Concerning viral infection we observe that the resulting solutions show local organization similar to that observed in cubic phases of amphiphilic molecules in water. Topologic comparison between bi-continuous liquid crystal phases and lamellar phases provides a guide for our study, as the geometric framework is the same, considering the interface between the amphiphilic molecule and the water as being structurally determining.

\subsection{Membrane mechanical properties and mechano-transduction}

Membrane mechanical properties, related to liquid crystal order and the fractal properties of the cytoskeleton, are driving the viscoelastic state of living tissue; they enable the organism to present solid elastic properties maintaining structural shape and viscous liquid properties modulating and optimizing interaction reaction times within biological processes as highlited by Julicher et al\cite{JulicherKruse}. An analogy could be the auditory perception of a galloping horse in the distance, which is transmitted faster and better on hard ground than sandy ground or through the air.

The tissues of living organisms combine elastic behaviors, related to solid like properties, in particular to maintain component morphology, and viscous liquid properties and energy sources (ATP hydrolysis) modulating global and local movements according to biological complexity. Energy balance have to take account of terms involved in classical elasticity, but also of dissipation related to viscosity (creation of entropy and  sources of energy).

Let us show how the SARS-CoV-2 hijacks the fractal properties of actomyosin and the liquid crystal state of the membrane containing domains which are modulated by the fractal consequences of the actomyosin contractility.

%
%
\begin{figure}[tbp]
\resizebox{1.0\textwidth}{!}{%
\includegraphics{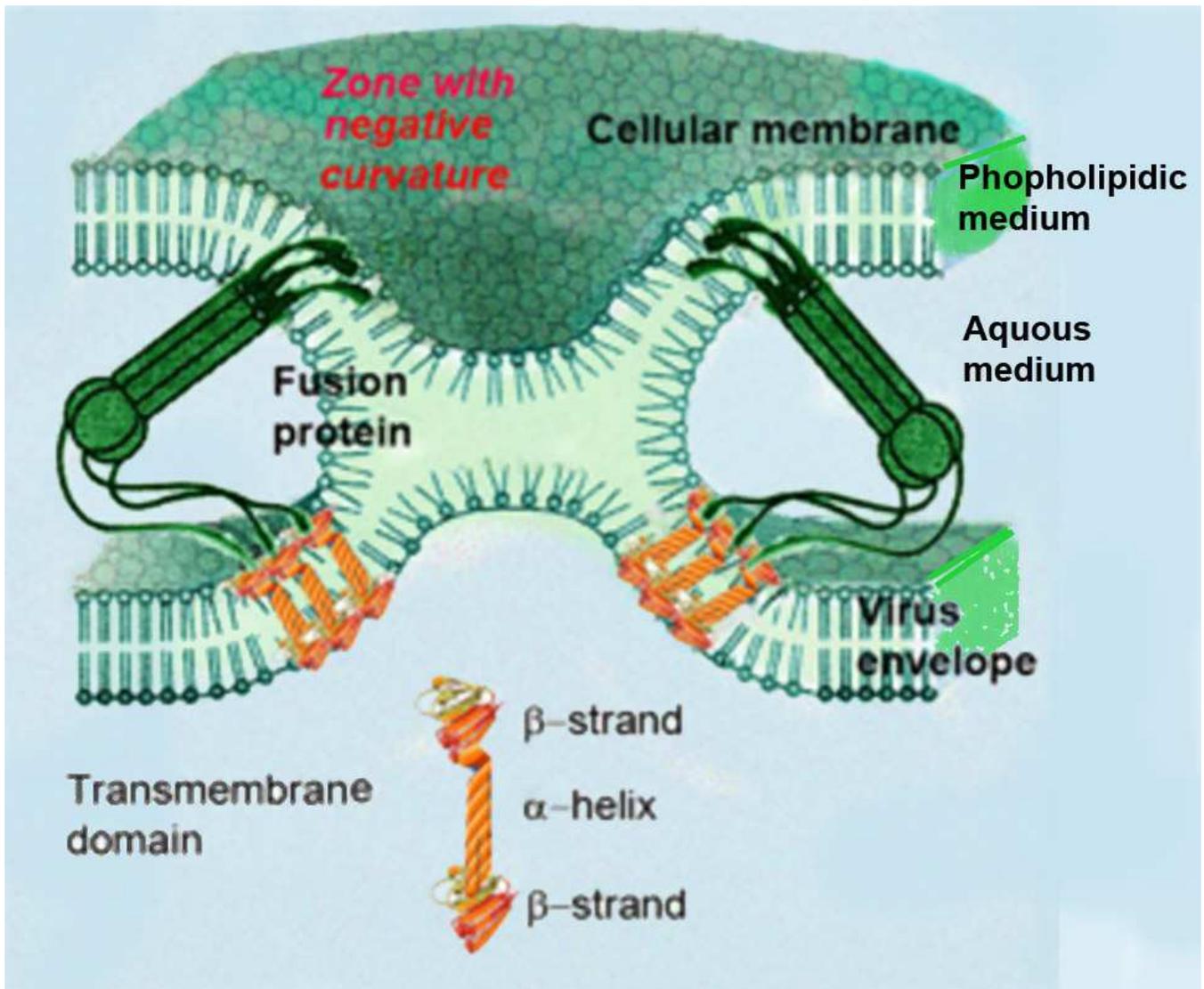} 
}
\caption{Fusion of membranes of viral and host cells. The Spike protein is the motor of the membranes fusion. It exhibits a $\beta$-strand-$\alpha$-helix-$\beta$-strand transmembrane conformation generating different membranes curvatures. This is associated with heterogeneity in phospholipid composition involving PE-rich domains.
Then a topological transformation involved by negative membrane curvature appears   as a hyperboloid-shaped tube which binds the two membranes.
This transformation related to diffusion of molecules leading to heterogeneity are created by change in the protein folding which provide the needed energy. Then three $\beta$-strands are immobilized by oligomerization. }
\label{f2}
\end{figure}
%
%

\section{SARS-CoV-2   hijacks  the fractal properties of actomyosin and membrane liquid crystal state}
\subsection{SARS-CoV-2  entry in target cell,  membrane fusion and mechano-transduction}

Coronaviruses, some influenza viruses and other pathogens enter the target cell within (TEMs) membrane tetraspanin-enriched microdomains. TEMs were identified as entry and exit portals for many viruses and, recently, as immunologic zones. These membrane domains extend the concept of lipid rafts that has been a focus of debate in recent decades. The SARS-CoV-2 receptor in the host cell plasma membrane is ACE2. SARS-CoV-2 also recognizes a transmembrane protein, neuropilin-1, to which it binds, facilitating interaction with ACE2. SARS-CoV-2 particularly activates a host cell membrane protease serin involved in viral envelope fusion with the host membrane: TMPRSS2\cite{MatsuyamaNao}, also located within TEMs. Upstream, other proteases, including furin, drive the proteolysis.
Spike protein binds to the ACE2 (hACE2) human membrane receptor. The Spike protein comprises two subunits: S1 initiates binding, and S2 is involved in membrane fusion.
Interaction between S1 and ACE2 decreases ACE2 activity. This activity downregulates the (RAAS) renin-angiotensin-aldosterone system, protecting the organism against its potentially harmful effects. This process can lead to many multi-organ dysfunctions, and respiratory complications in particular.

SARS-CoV-2 binds to ACE2 via a (RBD) receptor binding domain, which can be concealed from immune surveillance by a position that is mainly supine and less accessible than an upright position\cite{ShangWan}.

\subsection{SARS-CoV-2  and modification of molecular diffusion}
It is clear that virus entry modifies an existing mesophase or triggers a new mesophase. It has also been shown\cite{Sansonetti} that SARS-CoV-2 hijacks the activity of the cytoskeleton and its three components, including actomyosin, as soon as the virus binds to the target cell\cite{BassoVicente,Cherezov2003}. Virus entry thus modifies molecular diffusion and membrane molecule compartmentalization and hence dynamics, altering downstream cell signaling. Mechanical signaling is immediately involved in biochemical and genetic reaction cascades triggered by information relating the extracellular matrix to the intracytoplasmic nucleus, notably via cell adherences. We shall see that it is not surprising that the S1/ACE2 bond, for example, should destructure bronchial cell membranes, leading to lysis and respiratory depression followed by organ failure due to a massive, sometimes badly adapted immune response\cite{KubaImai,WangCaluch}. SARS-CoV-2 initiates and exploits a topological transformation in the membrane.

As early as 2009 it was known that many viruses, including enveloped RNA viruses, use various fusion proteins, inducing various mechanisms that change protein conformation and induce fusion, (corona, influenza, HIV1, Ebola, Dengue, HSV)\cite{RichardZhang}.

\subsection{SARS-CoV-2 and lamellar phase modification}
Phospholipid bilayer stability may depend on the ratio between the mean areas occupied by the polar heads (hydrophilic on the outer side of the membrane) and by the hydrophobic lipid chains within the membrane. In a planar membrane, as the cell membrane, the two are equal, as is also roughly the case for the viral membrane, at the cost of dissymmetry between the inner and outer sides. The infection mechanism involves the development of a tube (or tunnel) between the two membranes, in the form of a hyperboloid, figure~\ref{f1}  (i.e., with negative Gaussian curvature of surface) which is only stable if the polar heads occupy more lateral space than the hydrophobic chains\cite{CuiDeng}. This is a classic mechanism in phase transition from lamellar structures to “cubic” or “hexagonal” phases (see this book\cite{Gelbart} and in particular the chapter by Charvolin et al.). Such transformation is thus associated with changes in membrane curvature, and can be analyzed in terms of topologic change. In this transformation there is thus inevitably diffusion of membrane molecules, with properties differing from the mean, toward the contact area. The role of cholesterol as mediator is important here. The highjacking of the role of certain cell membrane proteins by SARS-CoV-2 very likely underlies this mechanism.

It should be stressed that these processes take place within the plasma membrane in most cases, or else in relation with the membrane via transmembrane proteins\cite{BarrettDutch}. Taking account of the liquid crystal state of the membrane phospholipid bilayer can shed light on early binding and membrane fusion processes and control of cytoskeletal domains with fractal properties modulating membrane nanodomains. In our view, the dipolar potential of the membrane bilayer is disturbed by the insertion of viral peptides, altering the membrane’s physical properties (dependence on lipid composition depending on the depth of peptide insertion in the membrane)\cite{TylerGreenfield,SimoesSilva}.

We consider that at this stage PE, within the membrane mesophases triggers the phase transition. Then, these disturbances underlie major complications related to SARS-CoV-2   infection. We argue this on the basis of the present state of knowledge, highlighting at each stage the role of membranes and actomyosin with their fractal properties. This links the first and the second part of the present paper, which concerns viral invasion.
Actomyosin may react to this rigidity, triggering a mechanical signal in response to the stress exerted by the virus. Membranes in good physiological condition, especially in young subjects, can limit viral invasion, but promote it in the elderly, whose membranes are impaired, notably by imbalance between polysaturated and insaturated lipids.

\subsection{SARS-CoV-2 initiates the bicontinuous phase}
The physics of phospholipid molecule structure\cite{CuiDeng,CharvolinSadoc2008} in the presence of water provides many examples of transition between lamellar and (often cubically symmetrical) bi-continuous phases. The term “bi-continuous” underlines the fact that these transitions pass from an environment divided up into a large number of phospholipid bi-layers separated by the corresponding number of water layers to just two continuous water environments separated by a single phospholipid film. This is a major topologic upheaval.

Fusion between cell and viral membranes in the critical stage of SARS-CoV-2   infection is driven energetically by passage from a metastable to a stable state of the Spike protein. Membrane fusion can thus be seem has  a lamellar-to-non-lamellar topologic transformation.
Interaction between the viral fusion peptide and the target cell membrane structure triggers a local topologic change by connecting the two membranes via a tunnel similar to those seen in bi-continuous phases. This topologic change enables communication between the cytoplasm and the genetic material of the virus.

In in-vitro experiments, B. Tenchov\cite{TenchovMacdonald} studied the mechanism of membrane fusion initiated by viral fusion peptides, using phospholipids of the DoPe-ME (dioleoylphosphatidylethanolamine
-N-methylethanalomine) family. Small-angle X-ray scattering (SAXS) showed that the N-terminal of the influenza virus hemagglutinin protein impacts formation of bi-continuous phases with lipid films of negative Gaussian curvature.

These results can be related to J. Charvolin and JF Sadoc studies showing that the geometrical parameters of membrane molecules and the energetic factors associated with interaction between bilayers govern the changes in curvature and distance between bilayers. These are the determining factors in the transformation from lamellar to cubic, hexagonal or micellar phase. Opposing stresses resolve in a compromise, bringing local curvatures into equilibrium\cite{CharvolinSadoc2011}.

Subunit S1 binding to the host-cell receptor has been shown to trigger a cascade of conformation changes in S2, initiating formation of an anti-parallel heteromeric bundle of six helices in the two heptad repeat regions. This process brings together the fusion peptide and the transmembrane domains. Now, the membrane interaction regions alter the physical properties of membranes that are composed of phospholipids. These processes related to membrane lipid composition depend on the depth of peptide insertion in the membrane, and thus its accessibility.

Peptide insertion disturbs the dipolar potential of the membrane bilayer, as mentioned above. Notice that in multilamellar phases observed in some liquid crystals the periodicity of layers in their orthogonal direction depends of the dipolar interaction through the water shell. It is also this dipolar interaction which is encountered here, allowing to virus and cell membrane  to come close together.

It is notable that, despite wide diversity of protein structure and nature, there is a shared process of fusion mediation enabling enveloped viruses to acquire a prior state suited for membrane fusion (dimer or trimer). Starting from this prior state, proteins adopt a structure that undergoes trimeric “hairpin” transformation, enabling the viral transmembrane domain to approach the fusion peptide bound to the target membrane\cite{KawaseKataoka}. Notice the crucial stage of protein transition from native metastable state to active stable state. The metastable native spike protein has an energy barrier preventing sudden change in conformation. But viral fusion peptides overpass this metastability leading to an energy transfer.

The glycoprotein subunit interacting with the host membrane contains two hydrophilic domains: a (FP) fusion peptide and a (TMD) transmembrane domain, that trimerize when associated by three  (figure-2)as described by Yao et al\cite{YaoLee}. The TMD includes a central helical nucleus, two extremities essentially composed of beta strands, and an N-terminal extremity that is more disordered than the rest of the peptide.
Close lipid membranes in water experience a strong mutual repulsion.  The TMD moderately increases the POPE (palmitoyloleoyl-phosphatidylethanolamine) membrane tendency to dehydrate leading to a decrease of the so-called hydration repulsion.

During cell/virus fusion, the primary protein is initially compact, then unfolds, exposing the fusion protein to the cell membrane and viral envelope. FP and TMD presumably induce formation of a fusion pore and completely fused membranes. Solid-state NMR shows that the influenza virus TMD fusion protein conformations are dependent on surrounding lipids. The beta-strand-rich conformation quantitatively transforms DOPE (1-2 dioleoyl-Sn-glycero-phosphatidylethanolamine) into some bi-continuous structure rich in negative saddle-like Gaussian curvature in the intermediate zones of hemi-fusion and fusion pores. Viral fusion protein TMD seems to mediate the necessary topologic changes in the membrane, the beta strand being used by the virus to achieve fusion.
Solid-state NMR and SAXS data provided the first structural proof, presented by Yao\cite{YaoLee}, that viral fusion TMD induces significant membrane curvature, especially in PE membrane. Beta-strand conformation correlates with curvature generation.

In all, a new model of viral fusion emerges, in which TMD promotes topologic changes in the membrane during fusion, using the beta strand as fusogenic conformation.

\subsection{SARS-CoV-2 and the innate immune system: involvement of mitochondria-associated membranes}

Mitochondria-associated membranes (MAM) are composed of membrane fragments from the endoplasmic reticulum and external mitochondrial membrane. Via a filament architecture, the two organelles establish contact areas forming a microdomain directly involved in immune response. The microdomain is a structure for scaffolding proteins and regulation factors. The contact sites lead to multiple involvement in human physiology and pathology\cite{BrisseLy}.
MAM form innate immune synapses, analogous to classical immune synapses. Below, we highlight a membrane stage that seems decisive in triggering host-cell immune response, involving (MAVS) mitochondrial antiviral signaling protein  on the external membrane of activated mitochondria.
PAMP (pathogen-associated molecular patterns) are detected by receptors recognizing specific host-cell patterns, including MDA5 (melanoma differentiation-associated protein 5), able to induce type-1 interferon. MDA5 is the receptor recognizing viral RNA. It recognizes viral RNA in the host-cell cytoplasm, and notably RNA containing a 5’-triphosphate with RNA double strand regions. MDA5 is an RLR (RIG-I-Like-RNA-helicase) able to bind to viral RNA\cite{ReikineNguyen}.
CARD (N-terminal caspase activation and recruitment domain) is a MDA5 domain inducing response to type-1 interferon. It interacts with the MAVS protein on the mitochondrial membrane via its C-terminal transmembrane domain. This mitochondrial location of MAVS is indispensable to trigger downstream antiviral signaling pathways. Activated MAVS induces recruitment of factors leading to phosphorylation of transcription factors. Several CARDs have to be aligned to allow interaction with MAVS, triggering downstream antiviral signaling. MAVS recruits ligases and facilitates activation of (IRF)  interferon regulation factors of (NF-k$\beta$) nuclear factor k$\beta$.
In a cell in absence of viral RNA, MDA5 CARDs are masked\cite{DiasSampaio} by intramolecular interactions with the helicase domain. After viral RNA binding, in contrast, a change in conformation mediated by ATP forms oligomers that expose their CARDs, which interact with the corresponding MAVS domain. The association of MDA5 with MAVS triggers formation of aggregates in mitochondrial surface which are characterized by their insolubility in detergents. This suggests the presence of nanodomains.   This conformation is the active state of MAVS described by Missiroli et al\cite{MissiroliPatergnani}.
In 2020, Hackbart\cite{HacbartDeng} highlighted a mechanism used by CoV endo-U to cleave viral RNA PAMP, modifying the recognition by MDA5 of the viral material and therefore altering the interaction of MDA5 with the MAVS protein. We propose that this mechanism lead to the early dysregulation of the innate immune response leading in particular to the discharge of cytokines.

Coronaviruses have an endoribonuclease, endo-U, which cleaves 5’polyuridine negative-sense viral RNA (PUN RNA). PUN RNA is a PAMP recognized by MDA5.

%
%
%
%

\subsection{SARS-CoV-2 and complications}

 We consider now pulmonary capillary endothelium functions and the involvement of intracellular membranes in the liquid crystal state, and cytoskeleton fractal properties.

It is established that cells sensitive to mechanical stresses activate signaling pathways transmitting information to the nucleus, generating or modifying cell response; these involve mechano-transduction.
Internal tension in the cytoskeleton related to mechano-transduction was identified as a crucial parameter in controlling vascular endothelium permeability in lung lesions\cite{WangCaluch}.
The barrier shows hyper-permeability, with strong elevation of circulating cells pass through the vascular compartment to the alveolar space, promoting inflammatory reaction.

There is a correlation between reinforcement of cell adhesion sites and increased substrate rigidity. Cytoskeleton remodeling redistributes intra- and inter-cellular adhesion, increasing permeability. Reinforced cell-matrix adhesion is accompanied by decreased cell-cell adhesion.
Loss of intercellular contact promotes edema, and release of cytokines, including IL1 and TNF-$\alpha$. These activate new neutrophils, altering the physical properties of the extracellular matrix. Fibroblasts colonize the interstitial space and produce matrix molecules, increasing extracellular matrix rigidity. At this stage, the plasma membrane itself is already structurally modified by the viral invasion, notably in terms of rigidity\cite{FavardChojnacki}.
As mentioned above, changes in the physical properties of the plasma membrane and intracellular membranes, including curvature, increased rigidity and altered viscoelasticity, generate the response of the fractal cytoskeleton, modifying molecular diffusion and hence cell signaling.
Loss of intracellular contact can be attributed to cytoskeleton remodeling, where fractal properties interact with the liquid crystal state of the membranes under mechanical stress. It would seem that SARS-CoV-2 invasion redistributes membrane phospholipids and that the underlying cytoskeleton modulates microdomain formation after SARS-CoV-2 binding to the ACE2 membrane receptor. Moreover, presence of TMPRSS2 membrane protein, involved in membrane fusion, suggests involvement of microdomains organized as tetraspanin platforms. These processes, we believe, increase the rigidity of the extracellular matrix and plasma membrane, altering molecular diffusion. Cytoskeleton actomyosin contractility increases internal tension, leading to a cell signaling cascade resulting in the respiratory distress observed in acute respiratory distress syndrome. Here we see a direct role of the liquid crystal state of pulmonary tissue mediated by fractal spaces generated by actomyosin contractility.
We would underline the role of pulmonary surfactant, which is a natural pure liquid crystal\cite{Mitov}. Surfactant integrity is a vital function, and these processes directly involve the liquid crystal state of the epi- and endo-thelial membranes and mediation by the contractile action of actomyosin with its fractal properties. It is the liquid crystal state that enables transition from a molecular unilayer on inspiration to a multilayer on expiration. In all, numerous chemical reactions, classically attributed to control of the pulmonary edema, seem to be themselves mediated by the mechanical properties of the extracellular matrix structure and by the mechano-transduction response of the cytoskeleton.

Controlling the major complication of SARS-CoV-2 thus involves taking account of and mastering parameters that are crucial to the mechanical equilibrium of the endothelial-alveolar-pulmonary barrier – whence the importance of determining, at cell level, the effects of mechanical stress on the cytoskeleton and its impact on endothelial function.

\section{Conclusion}

We implemented a multidisciplinary theoretical approach to processes involved in SARS-CoV-2 infection which highlight the modulation of cellular functions linked to membrane architecture and its alteration by SARS-CoV-2. Then we propose that actomyosin contractility initiates cytoskeletal fractal configurations. We know that the cytoskeletal controls the membrane domains and their mesophases\cite{Lehn,Honigmann} (or liquid crystal state), so most of the signaling from the extracellular matrix to the intracytoplasmic nucleus. It is accepted that viral disease modifies, and perverts, organization of the three components of the cytoskeleton, including its molecular motors.

By linking these questions to our previous research, we have  highlighted the involvement of the physical properties of membranes and the cytoskeleton under viral invasion and its impact on cell signaling. This is part of the study of our team for many years investigating physiological and pathological processes. One of the present authors has been studying liquid crystals for decades.

We now extend our exploration of pathogen invasion, using  concepts of mechano-transduction and far-from-equilibrium self-organization involved in energy transfers.
The form-function coupling, associated with the dynamics of structures includes mechanical stresses at the nanometric scale targeted by SARS-CoV-2. The viral infection therefore disrupts particularly molecular and ionic components diffusion, electrostatic effects and hydrogen bonding.

It is within mesomorphic states that the gene message and cellular communication are processed. Fusion peptides promote an organization in the membrane locally similar to inverse bi-continuous cubic structure. This topological transformation is notably under the influence of PE and cholesterol. In the self-organization of matter far from equilibrium and in mechano-transduction, mechanical information is inseparable from electrochemical and gene signaling, and has to be explored in viral infection.

There is an inseparable link between the, physical, chemical and genetic information of the cell; and its diversion by SARS-CoV-2.
For instance, an optimum reaction rate is required to select the formation of a trimer from among all potentially accessible solutions. The viscosity of the membranes, linked to the liquid crystal state of phospholipids, could promote speed control by biochemical catalysis. This leads to the trimerization of the fusion glycoprotein SPIKE. Then the trimer  provided the additional amount of free energy that a dimer could not provide  to cross the elastic deformation threshold of the membrane leading to the fusion pore.
This extra free energy overcomes the high kinetic barrier associated with hydrophobic repulsion in the aqueous intermembrane space when membranes come together below a few angstr\"oms apart. The n-terminus of the fusion peptide could induce dehydration of the inter-membrane space.
On the other hand, the negative curvature initiated by the fusion peptide and the oblique orientation of the latter disrupts the parallelism of the acylated lipid chains. The SARS-CoV-2 fusion peptide could thus promote the formation of a non-lamellar  phase at the bilayer level.

A lipid imbalance generated by the virus or pre-existing (due to age or an underlying disease) could promote the bicontinuous state coupling cell and virus membrane. Various studies\cite{ToelzerGupta} mention and explore processes concerning, for example, essential free fatty acids such as linoleic acid during viral binding to the membrane receptors of the host cell ACE2. These intramembrane processes are  linked to nanodomains. Notice that the membrane lipid imbalance has also been documented in the field of cancerology. The latter is increasingly understood as a signaling disease\cite{ChouardBinotHelyon}. The same is true for many neurodegenerative pathologies and prionic conditions. The question, for example, of the transmissibility of Alzheimer's disease\cite{Dhenain2021} is now being asked. We propose to link these questions to supramolecular shape memory involving membrane mesophases\cite{PetercaImam}, Van der Waals bonds, hydrogen bond, ionic forces, beyond intramolecular covalences.

Taking into account the physical properties of cells, hijacked by SARS-CoV-2, broadens the exploration of the mechanisms leading to COVID-19.

We have linked the data from the literature concerning geometry and cell topology, beyond all the well accepted works concerning membrane signalization modulated by the underlying cytoskeleton.

We state, from recent investigations, that the actomyosin contractility of the cytoskeleton generates structures showing fractal properties. It is observed that the cytoskeleton modulates the membrane nanodomains participating to the processes of cell signaling. These mechanisms contribute to the transmition of information from the extracellular matrix to the intracytoplasmic nucleus. The virus (SARS-CoV-2 and others) is shown to take control of this chain of transmission.

Consequently, controlling the dynamic contractility of the actomyosin is a global purpose in the fight against viral invasion. It is part of a multidisciplinary approach involving all scales of distance and time in living organisms, particularly in the field of the self-organization of matter out of static equilibrium (and so consuming the ATP of its constituents).

In this context, we have investigated the fusion of viral and host cell membranes as a local mechanism. It is assumed that the membrane phospholipid bilayer is in a liquid crystal or mesophase state.
We emphasize on the role of the SARS-CoV-2 fusion peptide which changes the lamellar organisation and transforms it (in the presence of ethanolamine) into a bicontinuous conformation involving topological modifications.

We insist on the fact that a common information vector linked to membrane signaling involves chemistry, physics and genetics which have to be strongly imbricated. It could be involved in infection by pathogens, neurodegeneration, oncogenic transformation, and more generally to aging of the body, within a generic mechanism common to all of these processes.

We have emphasized on the structural role of cell membrane, but it is also important to underline the involvement of mitochondrial membranes in the innate immune response deregulated by SARS-CoV-2. This virus has an endoribonuclease interfering with the cellular response in the presence of the viral genetic material.
We propose that the disrupted interaction between cytoplasmic MDA5, viral material, and MAS protein, localized on the mitochondrial outer membrane, is a crucial step in the deregulation of the innate immune response. A therapeutic approach targeted at the MAM area could participate in the control of the inappropriate mechanism of the host cell response to viral invasion.

Geometry and topology are central in processes related to COVID19. Films of amphiphilic molecules organize themselves between local molecular behavior and the overall shape of membranes. This results in long-distance interactions favoring the control by the virus of fundamental cellular processes as soon as it binds to the membrane receptor ACE2.

The fight against SARS-CoV-2 involves taking into account the physical properties of cells and particularly the geometry and membrane topology.
It is by using all the fundamental knowledge available, including   the notions of order, and disorder, and those of mathematical symmetries that we will progress as quickly as possible towards useful "bedside" solutions.

The pandemic prompts us to overcome multidisciplinary obstacles that fundamental research has come up against for several decades.


\end{document}